\documentclass[preprintnumbers, nofootinbib,amsmath,prd,aps,superscriptaddress,tightenlines,12pt]{revtex4}
\usepackage[hypertex]{hyperref}
\usepackage{graphicx}
\usepackage{epstopdf}
\usepackage{bm}
\usepackage{epsfig}
\usepackage{graphics}
\usepackage{xspace}
\usepackage{pstricks}
\usepackage{subfigure}
\usepackage{psfrag}

\def\nslash{n\!\!\!\slash}

\def\bnslash{\bar n\!\!\!\slash}

\def\OMIT#1{}

\newcommand{\nn}{\nonumber}

\newcommand{\bn}{{\bar n}}
\newcommand{\bea}{\begin{eqnarray}}
\newcommand{\eea}{\end{eqnarray}}

\newcommand{\be}{\begin{equation}}
\newcommand{\ee}{\end{equation}}

\begin{document}
\setlength\baselineskip{17pt}
\setlength{\unitlength}{1cm}



\preprint{NPAC-10-06}

\title{\bf Parity-Violating Electron-Deuteron Scattering with a Twist}

\author{Sonny Mantry}
\email[]{mantry147@gmail.com}
\affiliation{University of Wisconsin-Madison, Madison, WI 53706}
\author{Michael J. Ramsey-Musolf}
\email[]{mjrm@physics.wisc.edu}
\affiliation{University of Wisconsin-Madison, Madison, WI 53706}
\affiliation{California Institute of Technology, Pasadena, CA 91125}
\author{Gian Franco Sacco}
\email[]{gianfranco.sacco@jpl.nasa.gov}
\affiliation{Jet Propulsion Lab/ California Institute of Technology,
   Pasadena, CA, 91109}



\newpage
\begin{abstract}
  \vspace*{0.3cm}
We show that Parity-Violating Deep Inelastic Scattering (PVDIS) of longitudinally polarized electrons from deuterium can in principle be a relatively clean probe of  higher twist quark-quark correlations beyond the parton model. As first observed by Bjorken and Wolfenstein, the dominant contribution to the electron polarization asymmetry, proportional to the axial vector electron coupling, receives corrections at twist-four from the matrix element of a single four-quark operator. We reformulate the Bjorken/Wolfenstein argument in a matter suitable for the interpretation of experiments planned at the Thomas Jefferson National Accelerator Facility (JLab). In particular, we observe that because the contribution of the relevant twist four operator satisfies the Callan-Gross relation, the ratio of parity-violating longitudinal and transverse cross sections, $R^{\gamma Z}$, is identical to that for purely electromagnetic scattering, $R^{\gamma}$,  up to perturbative and power suppressed contributions. 
This result simplifies the interpretation of the asymmetry in terms of other possible novel hadronic and electroweak contributions. We use the results of MIT Bag Model calculations to estimate contributions of the relevant twist four operator to the leading term in the asymmetry as a function of Bjorken $x$ and $Q^2$. We compare these estimates with possible leading twist corrections due to violation of charge symmetry in the parton distribution functions. 

 \end{abstract}

\maketitle

\newpage

\section{Introduction}

In the 1970s parity-violating deep inelastic scattering (PVDIS) of longitudinally polarized electrons from deuterium played an important role in confirming the Standard Model (SM) of particle physics~\cite{Prescott:1978tm, Prescott:1979dh, Cahn:1977uu}.  The asymmetry
\bea
A_{RL} &=& \frac{\sigma_R - \sigma_L}{\sigma_R + \sigma_L},
\eea
with $\sigma_{R,L}$ corresponding to the scattering cross-section with positive and negative helicity electrons respectively, is an excellent probe of the parity-violating electroweak interactions in the SM. The results of measuring this asymmetry in the early experiments at SLAC led to the correct description of neutral weak interactions well before the discovery of the $Z$ boson at CERN and provided a measurement of the weak mixing angle $\sin ^2 \theta_W$. Since then parity-violating electron scattering  from various targets has been studied at JLab~\cite{Armstrong:2005hs,Acha:2006my,Aniol:2007zz,:2009zu}, MIT/Bates~\cite{Ito:2003mr,Spayde:2003nr}, Mainz~\cite{Heil:1989dz,Baunack:2009gy}, and SLAC\cite{Anthony:2005pm}
as a tool for probing physics beyond the SM and hadronic structure. 

Currently, an active program is underway at JLab to continue these studies with a new level of precision. The Q-Weak experiment ~\cite{VanOers:2007zz}, which will measure the weak charge of the proton at low  electron momentum transfer ($Q^2$), is expected to determine $\sin^2\theta_W$ to  $0.3\%$ precision, making it the most precise test of the running of the weak mixing angle to date. Furthermore, the recently approved 12 GeV upgrade of CEBAF at JLab, expected to be completed by 2014, aims to begin the next generation Moller  and electron-deuteron scattering experiments. The SOLID proposal~\cite{Souder:2008zz} for precision parity-violating  electron-deuteron scattering, approved as part of the 12 GeV upgrade, will measure  $A_{RL}$ over a wide kinematic range in $Q^2$ and Bjorken-$x$ to within $1\%$  at each kinematic point.  In addition, one high-precision PVDIS experiment with deuterium has completed data taking at selected kinematic points with the 6 GeV \cite{PVDIS:Jlab6} beam and another is approved to run at the 12 GeV \cite{PVDIS:JLab12} beam. These present and prospective  high-precision experimental measurements present new challenges for their theoretical interpretation.
In particular, substantial uncertainties in the theoretical interpretation of the deep inelastic asymmetries will remain unless various effects contributing to the asymmetry such as new physics beyond the SM, sea quark distributions, Charge Symmetry Violation (CSV), and higher twist contributions are well understood and disentangled from each other. Addressing one aspect of these issues is the subject of this paper.

The theoretical interpretation of the deuterium asymmetry can be facilitated by expressing it in the following form
\bea
\label{APVY1Y3-A}
A_{RL} &=& - \Bigg ( \frac{G_F Q^2}{4\sqrt{2} \pi \alpha} \Bigg )  \Bigg [ g_A^e Y_1 \frac{F_1^{\gamma Z}}{F_1^\gamma} + \frac{g_V^e}{2} Y_3 \frac{F_3^{\gamma Z}}{F_1 ^\gamma}\Bigg ].
\eea
Here, $g_V^e$ ($g_A^e$) are the vector (axial vector) couplings of the Z-boson to the electron; $F_1^\gamma$,  $F_1^{\gamma Z}$, and $F_3^{\gamma Z}$ are the structure functions arising, respectively, from hadronic matrix elements of the vector electromagnetic (EM) current, interference of the vector EM and vector weak neutral current (WNC), and interference of the vector EM current and axial vector WNC; and $Y_{1,3}$ are functions of kinematic variables and the ratios $R^\gamma$ and $R^{\gamma Z}$ of longitudinal and transverse cross sections for purely EM and WNC-EM vector current interference cross sections.  In the SM, at leading twist and in the absence of CSV effects, the $Y_1$ term in Eq.(\ref{APVY1Y3-A}) is independent of $y$ and depends only on $g_A^e$ and the vector current coupling of the $Z$-boson to quarks~\cite{Cahn:1977uu}. Since $g_V^e=-1+4\sin^2\theta_W\sim -0.1$, the $Y_1$-term dominates the asymmetry, making its scrutiny particularly important for the interpretation of the Jefferson Lab PVDIS program.

Considerable theoretical effort has been devoted to disentangling the various contributions to the asymmetry. The effect of twist-four contributions to the asymmetry was first considered in papers by Bjorken and Wolfenstein~\cite{Bjorken:1978ry,Wolfenstein:1978rr} more than thirty years ago, where it was shown to arise in the dominant axial electron coupling term from a single, non-local four-quark operator in the limit of good isospin, negligible sea-quark and CSV effects, and up to corrections in $\alpha_s(Q^2)$. Quantitative estimates of twist-four effects were first obtained in
 \cite{Fajfer:1984um} where the contribution of the spin-two operators was estimated using the MIT Bag Model. This analysis was extended in \cite{Castorina:1985uw} to include corrections to the $F_3^{\gamma Z}$ structure function. More recently, higher twist effects to the asymmetry were estimated by the authors of Ref.~\cite{Hobbs:2008mm}, who considered the possibility that  $R^{\gamma}$ and $ R^{\gamma Z}$  receive substantially different contributions from finite-$Q^2$ effects. These authors argued that such a difference could introduce hadronic uncertainties that might impede the extraction of CSV effects from $A_{RL}$.
 
In this paper, we draw on the observations of~\cite{Bjorken:1978ry,Wolfenstein:1978rr} that the twist-four contribution to the $Y_1$ term in $A_{RL}$  for deuterium, given in Eq. (\ref{APVY1Y3-A}), arises from a single four-quark operator involving up- and down-quark fields
\be
\label{Omunu}
\mathcal{O}^{\mu\nu}_{ud}(x) = \frac{1}{2} [{\bar u}(x)\gamma^\mu u(x) d(0)\gamma^\nu d(0) + (u\leftrightarrow d) ] \ \ \ 
\ee
to revisit the analysis of Ref.~\cite{Hobbs:2008mm}. Noting that the contribution of ${O}^{\mu\nu}_{ud}(x)$ to the electroweak structure functions satisfies the Callan-Gross relation at leading order in the strong coupling,  we find that 
\bea
R^{\gamma Z} &=& R^\gamma \qquad \text{and} \qquad Y_1=1,
\eea
at twist-four up to perturbative corrections. Consequently, all  twist-four effects entering the dominant term in the asymmetry reside in the ratio $
F_1^{\gamma Z}/F_1^\gamma$ . 

Using the power law dependence in $Q^2$ of the twist-four effects to the $Y_1$-term it may be possible, with the precision and the wide kinematic range of the PVDIS program at JLab and its possible extension at an electron-ion collider~\cite{Kumar:2010} , to disentangle twist-four effects from CSV effects depending on their relative overall sizes. To provide theoretical guidance for such a program, we utilize the MIT Bag Model\cite{Chodos:1974je} to estimate the size and variation of the  twist-four contribution with Bjorken-$x$ and $Q^2$.   These estimates extend the earlier work of Ref.~\cite{Castorina:1985uw} by allowing for the $x$-dependences of the twist-two and twist-four contributions to $F_1^{\gamma(\gamma Z)}$ to differ. We find that if the MIT Bag Model reasonably estimates the magnitude of the  twist-four contribution from ${O}^{\mu\nu}_{ud}(x)$, the impact on the asymmetry would likely be too small to be extracted  without further improvements in experimental precision. In this case, however, the planned PVDIS experiments could in principle provide a theoretically clean probe of possible contributions from CSV and/or physics beyond the SM. Conversely, the observation of significant power corrections to the $Y_1$ term would signal the presence of relatively large and theoretically interesting quark-quark correlation contributions to the electroweak structure functions.

Our analysis leading to these conclusions is organized in the remainder of the paper as follows. In Section \ref{sec:over},  we provide an overview of the structure of the deuterium asymmetry, setting the context for our analysis of the twist-four contributions in Section \ref{sec:twistfour}. In Section \ref{sec:mit} we give our MIT Bag Model estimates and compare these with recent parameterizations of CSV contributions as well as possible effects from ``new physics" in Section \ref{sec:csv}. We summarize our conclusions in Section \ref{sec:conclude}. In appendix ~\ref{cgdu}, we also recast the argument of~\cite{Ellis:1982cd, Ji:1993ey,Qiu:1988dn}  in the language of the Soft-Collinear Effective Theory(SCET)~\cite{Bauer:2000yr, Bauer:2001yt, Bauer:2002nz} that shows manifestly that the twist-four matrix element contributing to the $Y_1$-term satisfies the Callan-Gross relation at tree level in the matching.

\section{Overview}
\label{sec:over}

Before presenting the formalism and derivation of our results, we provide an overview of the structure of the deuterium asymmetry and the context for the higher twist contributions. The SM parity violating interactions of the electron with the quarks, obtained after integrating out the Z-boson, are parameterized as 
\bea
\label{ele-quark}
{\cal L} &=& \frac{G_F}{\sqrt{2}} \Big [ \bar{e}\gamma^\mu \gamma_5 e \big (C_{1u} \bar{u} \gamma_\mu u + C_{1d} \bar{d} \gamma_\mu d \big ) + \bar{e}\gamma^\mu e \big ( C_{2u} \bar{u} \gamma_\mu \gamma_5 u + C_{2d} \bar{d} \gamma_\mu \gamma_5 d \big ) \Big ], 
\eea
where  the coefficients $C_{1q}$ and $C_{1q}$ are given by
\bea
\label{C12rad}
C_{1q} & = & 2 {\hat\rho}_{NC} I_3^e \left( I_3^q-2 Q_q {\hat\kappa}\sin^2{\hat\theta}_W\right) -\frac{1}{2}{\hat\lambda}_1^q\\
C_{2q} & = & 2 {\hat\rho}_{NC} I_3^q \left( I_3^e-2 Q_e {\hat\kappa}\sin^2{\hat\theta}_W\right) -\frac{1}{2}{\hat\lambda}_2^q\ \ \ .
\eea
Here $I_3^f$ is the third component of weak isospin for fermion $f$, $Q_f$ is the electromagnetic charge, and ${\hat\theta}_W$ is the weak mixing in the $\overline{\mathrm{MS}}$ scheme. The quantities $ {\hat\rho}_{NC}$, ${\hat\kappa}$, and ${\hat\lambda}_j^q$ encode the effects of electroweak radiative corrections and take on the values one, one, and zero, respectively, at tree-level, leading to
\bea
\label{C12}
C_{1u}^\mathrm{tree} &=& -\frac{1}{2} + \frac{4}{3} \sin^2 \theta_W, \qquad C_{1d}^\mathrm{tree} = \frac{1}{2} - \frac{2}{3}\sin^2\theta_W, \nn \\
C_{2u}^\mathrm{tree} &=&  -\frac{1}{2} + 2\sin^2 \theta_W, \qquad C_{2d}^\mathrm{tree} =  \frac{1}{2} -2 \sin^2 \theta_W\ \ .
\eea

The reason for the high sensitivity of $A_{RL}$ to these interactions is that in the limit of good isospin and negligible sea quark effects, all hadronic effects are known to cancel in the asymmetry  at leading order in the twist expansion (corresponding to the parton model limit). The resulting expression for the asymmetry, known as the Cahn-Gilman  (CG) formula~\cite{Cahn:1977uu},  is given at tree-level by\footnote{We observe that the RHS of Eq.(4.89) of Ref. \cite{Musolf:1993tb} should be multiplied by (-1).}
\bea
\label{cahn}
A^{RL}_{\text{CG}} &=& -\frac{G_F Q^2}{2\sqrt{2}\pi \alpha} \frac{9}{10}\Big [ \big (1-\frac{20}{9}\sin^2\theta_W \big ) + \big (1- 4 \sin^2 \theta_W \big ) \frac{1-(1-y)^2}{1+(1-y)^2}\Big ]\ \ \ .
\eea
Here $y$ is the kinematic variable defined  as
\bea
\label{y-var}
y &=& \frac{2 P\cdot (\ell - \ell')}{2 P \cdot \ell},
\eea
where $P_\mu$, $\ell_\mu$, and $\ell_\mu'$ denote the four momenta of the deuteron, the incoming electron, and the outgoing electron respectively. In the lab frame, one has $y= (E-E')/E$ where $E$ and $E'$ denote of the energies of the incoming and and outgoing electrons. The corrections to this Cahn-Gilman formula can be parameterized by writing the asymmetry as
\bea
\label{CG-modify}
A_{RL} &=& - \frac{G_F Q^2}{2\sqrt{2}\pi \alpha}\frac{9}{10}\Big [\tilde{a}_1  + \tilde{a}_2 \frac{1-(1-y)^2}{1+(1-y)^2}\Big ],
\eea
where the parameters $\tilde{a}_{j}$ ($j=1,2$) are schematically written as
\bea
\label{R-effects}
\tilde{a}_j &=& -\frac{2}{3}\left(2C_{ju}-C_{jd}\right) \big [1 + R_j(\mathrm{new})  +  R_j(\mathrm{sea})+ R_j(\mathrm{CSV})+ R_j(\mathrm{TMC}) + R_j(\mathrm{HT})  \big ]
\eea
and $R_{j}(\mathrm{new})$, $R_{j}(\mathrm{sea})$, $R_{j}(\mathrm{CSV})$, $R_{j}(\mathrm{TMC})$, and $R_{j}(\mathrm{HT})$ denote respectively corrections arising from possible new physics beyond the SM,  sea quark effects, CSV, target mass corrections (TMC)~\cite{Schienbein:2007gr, Accardi:2008ne}, and higher twist (HT) contributions.  If one is interested in looking for signals of new physics beyond the SM that can leave a footprint in the asymmetry via the contributions $R_{1,2}(\mathrm{new})$, it is crucial that all the SM electroweak and hadronic corrections to the Cahn-Gilman formula in Eq.~(\ref{R-effects}) are under theoretical and experimental control. One can take an alternative viewpoint and instead view a precision measurement of $A_{RL}$ as a probe of hadronic physics that modifies the Cahn-Gilman formula as in Eqs.(\ref{CG-modify}) and (\ref{R-effects}).  

The analysis of this paper is focused on the higher twist correction $R_1(\mathrm{HT})$ that enters the $\tilde{a}_1$ term of the asymmetry. The leading contribution to $R_1(\mathrm{HT})$ appears at twist-four, giving rise to a $1/Q^2$ power law dependence. In contrast, the leading contribution from $R_1(\mathrm{TMC})$, which will also have a $1/Q^2$ power law contribution, will be suppressed relative to $R_1(\mathrm{HT})$.  The relative suppression of $R_1(\mathrm{TMC})$ can be understood by noting that the derivation of the Cahn-Gilman formula is valid even for a finite target mass so that target mass corrections will always appear in conjunction with at least one of the already small effects that correct the Cahn-Gilman formula. 

Given that all the remaining contributions to $\tilde{a}_1$ in Eq.~(\ref{R-effects}) have at most a logarithmic dependence on $Q^2$, one can, in principle, make a clean extraction of $R_1(\mathrm{HT})$ by studing the $Q^2$ dependence of the $\tilde{a}_1$ term in the asymmetry. Similar statements can be made for the higher twist effects $R_2(\mathrm{HT})$ that contribute to the $\tilde{a}_2$ term of the asymmetry. However, the study of $R_1(\mathrm{HT})$  is particularly interesting because  the leading contribution to $R_1(\mathrm{HT})$ that arises at twist-four is given entirely by a single matrix element that characterizes quark-quark correlations in the deuteron as first observed in ~\cite{Bjorken:1978ry,Wolfenstein:1978rr}. This is in contrast to $R_2(\mathrm{HT})$, which receives contributions from several different twist-four matrix elements making it difficult to interpret the effect of correlations among quarks and gluons in terms of any one of the these matrix elements. 

Before giving the explicit expression for $R_1(\mathrm{HT})$ that we derive below, we first review some of the standard notation used in PVDIS pehenomenology. The  general expression for the  asymmetry $A_{RL}$ is given in terms of the five structure functions $F_{1,2}^\gamma$ and $F_{1,2,3}^{\gamma Z}$ takes the form \cite{Hobbs:2008mm}
\bea
\label{asym1}
A_{RL} &=& - \Big ( \frac{G_F Q^2}{4\sqrt{2} \pi \alpha} \Big ) \frac{g_A^e \big (2 xy F_1^{\gamma Z} - 2 \big [ 1- 1/y + \frac{xM}{E} \big ] F_2^{\gamma Z} \big) + g_V^e x (2-y)F_3^{\gamma Z}}{2xy F_1^\gamma - 2 \big [ 1- 1/y + \frac{x M}{E}\big ]F_2^\gamma}.
\eea
This general expression reduces to the Cahn-Gilman formula  when the leading twist and isospin limits are applied to structure functions and when sea quark  and CSV effects are ignored. The $F_{1,2}^{\gamma Z}$ and $F_{3}^{\gamma Z}$ structure functions arise from the interference of the electromagnetic current with the vector and axial part of the weak neutral current respectively. 
The asymmetry is often parameterized in terms of the ratio of the longitudinal to transverse virtual neutral vector boson cross-sections 
\bea
\label{R-gamma-Z}
R^{\gamma (\gamma Z)} \equiv \frac{\sigma_L^{\gamma (\gamma Z)}}{\sigma_T^{\gamma (\gamma Z)}} =r^2 \frac{F_2^{\gamma(\gamma Z)}}{2xF_1^{\gamma(\gamma Z)}} -1, \qquad r^2 = 1+ \frac{4M^2 x^2}{Q^2}.
\eea
In terms of $R^{\gamma (\gamma Z)}$ the asymmetry in Eq.~(\ref{asym1}) takes the form given in Eq.~(\ref{APVY1Y3-A}), 
where the quantities $Y_{1,3}$ are defined as
\bea
\label{y1andy3}
Y_1 &=& \Bigg ( \frac{1+R^{\gamma Z}}{1+ R^\gamma}\Bigg ) \> \frac{1 + (1-y)^2 - y^2\Big [1-r^2/(1+R^{\gamma Z})\Big] - 2xy M/E}{1 + (1-y)^2 - y^2\Big [1-r^2/(1+R^{\gamma })\Big] - 2xy M/E}, \nn \\
Y_3 &=& \Bigg ( \frac{r^2}{1+ R^\gamma}\Bigg ) \> \frac{1 - (1-y)^2 }{1 + (1-y)^2 - y^2\Big [1-r^2/(1+R^{\gamma })\Big] - 2xy M/E}. \nn \\
\eea
In this notation, the $Y_1$ and $Y_3$ terms arise from the interference of the electromagnetic current with the vector and axial-vector weak neutal current respectively.

One of the main results of this paper is that the relation
\bea
R^{\gamma} &=& R^{\gamma Z} = r^2-1,
\eea
known to hold at leading twist due to the Callan-Gross relations of the structure functions, also holds  even after the twist-four contributions to $R_1(\mathrm{HT})$ are included at tree level. Equivalently, the relation
\bea
Y_1=1,
\eea
is valid at twist-four up to perturbative corrections in $\alpha_s(Q^2)$.  However, the the twist-four contribution does affect the ratio $F_1^{\gamma Z}/F_{1}^\gamma$ in the $Y_1$-term of Eq.~(\ref{APVY1Y3-A}) as
\bea
\label{R1HTA}
\Bigg [\frac{F_1^{\gamma Z}}{F_1^\gamma} \Bigg ]_{\text{CG + HT}}&=& -\frac{6}{5}(2C_{1u}-C_{1d})\Big [ 1 + R_1(HT)  \Big ]\\
\label{F-ratio-intro}
&=& \frac{9}{5}(1-\frac{20}{9}\sin^2 \theta_W)\Big [ 1 + R_1(\mathrm{HT})  \Big ],  \nn
\eea
where the first term corresponds to the Cahn-Gilman limit and where, in the second line we have omitted the electroweak radiative corrections for simplicity of presentation as we will do throughout much of the remainder of the paper.

\section{Isolating the twist-four contribution}
\label{sec:twistfour}

\subsection{Structure Functions}

In this section we review the basic phenomenology and conventions for electron-deuteron PVDIS. The differential cross-section for electron-deuteron scattering takes the general form
\bea
\label{diff-cross}
\frac{d^2\sigma}{d\Omega dE'} = \frac{\alpha^2}{Q^4}\frac{E'}{E} \Big ( L_{\mu \nu}^\gamma W_\gamma^{\mu \nu} - \frac{G_FQ^2}{4\sqrt{2} \pi \alpha} L_{\mu \nu}^{\gamma Z}W_{\gamma Z}^{\mu \nu}   \Big ),
\eea
where $E$ and $E'$ denote  the energies of the incoming  and outgoing electron respectively in the lab frame.  The square of the momentum transfer via the exchanged photon or Z-boson is $Q^2=-q^2=-(\ell -\ell ')^2$ where $\ell_\mu $ and $\ell_\mu'$ denote the four-momenta of the incoming and outgoing electron respectively.  The leptonic tensors in Eq.~(\ref{diff-cross}) are given by
\bea
L_{\mu \nu}^\gamma &=&  2 (\ell_\mu \ell_\nu' + \ell'_\mu \ell_\nu - \ell \cdot \ell' g_{\mu \nu}- i \lambda \epsilon_{\mu \nu \alpha \beta} \ell^\alpha \ell^{'\beta} ), \nn \\
L_{\mu \nu}^{\gamma Z}&=& (g_V^e + \lambda g_A^e)L_{\mu \nu}^\gamma,
\eea
where $\lambda$ denotes the sign of the initial electron helicity with $\lambda=1, -1$ for positive and negative helicity states respectively. The hadronic tensors in Eq.~(\ref{diff-cross}) take the form
\bea
\label{W-def}
W_{\mu \nu}^{\gamma (\gamma Z)} &=& \frac{1}{2M} \sum_{X} (2\pi)^3 \delta^{(4)} (p_X - P -q)\nn \\
&\times& \Big \{ \langle X | J_{\mu}^{\gamma (Z)} | D(P) \rangle ^*\langle X |J_\nu ^\gamma | D(P) \rangle  + \langle X | J_{\mu}^{\gamma } | D(P) \rangle ^*\langle X |J_\nu ^{\gamma (Z)}| D(P) \rangle \Big \},
\eea
where $J_\mu^{\gamma}$  and $J_\mu^{Z}$ denote the quark current coupling to the exchanged photon and Z-boson respectively and $M$ denotes the deuteron mass. The hadronic tensors are parameterized in terms of the structure functions $F_{1,2}^\gamma$ and $F_{1,2,3}^{\gamma Z}$ as
\bea
W_{\mu \nu}^{\gamma} &=& \big ( -g_{\mu \nu} + \frac{q_\mu q_\nu}{q^2}\big)\frac{F_1^\gamma}{M} + \big(P_\mu - \frac{P\cdot q}{q^2} q_\mu \big)\big (P_\nu - \frac{P\cdot q}{q^2}q_\nu \big)\frac{F_2^\gamma}{ M P\cdot q}, \nn \\
W_{\mu \nu}^{\gamma Z} &=& \big ( -g_{\mu \nu} + \frac{q_\mu q_\nu}{q^2}\big )\frac{F_1^{\gamma Z}}{M} + \big (P_\mu - \frac{P\cdot q}{q^2} q_\mu \big) \big (P_\nu - \frac{P\cdot q}{q^2}q_\nu \big )\frac{F_2^{\gamma Z}}{ M P\cdot q}+ \frac{i \epsilon_{\mu \nu \alpha \beta} P^\alpha q^\beta}{2 M P \cdot q} F_3^{\gamma Z}. \nn \\
\eea
The structure functions depend on two variables conventionally taken to be $Q^2$ and  Bjorken  $x = Q^2/(2 P\cdot q)$.
The definitions of the $F_{1,2,3}^{\gamma Z}$ structure functions in terms of the vector and axial vector neutral weak current operators can be obtained by first breaking up the weak neutral current into its vector ($J^Z_{V\mu}$) and axial-vector ($J^Z_{A\mu}$)  parts so that
\bea
J^Z_\mu = J^Z_{V\mu} + J^Z_{A\mu},
\eea
which allows a decomposition of the hadronic tensor $W^{\gamma Z}_{\alpha \beta}$ in Eq.~(\ref{W-def}) as
\bea
W_{\alpha \beta}^{\gamma Z} = W_{\alpha \beta}^{V; \gamma Z} +  W_{\alpha \beta}^{A; \gamma Z},
\eea
where $W_{\alpha \beta}^{V(A); \gamma Z}$ correspond to the hadronic tensors arising from the interference of the electromagnetic current with the vector and axial-vector weak neutral current respectively and are given by
\bea
\label{WVA-1}
W_{\mu \nu}^{V(A); \gamma Z} &=& \frac{1}{2M} \sum_{X} (2\pi)^3 \delta^{(4)} (p_X - P -q)\nn \\
&\times& \Big \{ \langle X | J_{V(A)\mu}^{Z} | D(P) \rangle ^*\langle X |J_\nu ^\gamma | D(P) \rangle  + \langle X | J_{\mu}^{\gamma } | D(P) \rangle ^*\langle X |J_{V(A)\nu} ^{Z}| D(P) \rangle \Big \}.\nn \\
\eea
In terms of the above hadronic tensors the  $F_{1,2,3}^{\gamma Z} $ structure functions are  given by 
\bea
W_{\mu \nu}^{V; \gamma Z} &=& \big ( -g_{\mu \nu} + \frac{q_\mu q_\nu}{q^2}\big)\frac{F_1^{\gamma Z}}{M} + \big(P_\mu - \frac{P\cdot q}{q^2} q_\mu \big)\big (P_\nu - \frac{P\cdot q}{q^2}q_\nu \big)\frac{F_2^{\gamma Z}}{ M P\cdot q}, \nn \\
W_{\mu \nu}^{A;\gamma Z} &=&  \frac{i \epsilon_{\mu \nu \alpha \beta} P^\alpha q^\beta}{2 M P \cdot q} F_3^{\gamma Z}. \nn \\
\eea

\subsection{Isospin decomposition of structure functions}

We now show in the limit of good  isospin and negligible sea quark contributions,  the twist-four contributions to the $Y_1$ term in Eq. (\ref{APVY1Y3-A}) come purely from four-quark twist-four operators up to possible higher order perturbative mixing effects involving quark-gluon or purely gluonic operators. This result was first pointed out in
 \cite{Bjorken:1978ry,Wolfenstein:1978rr}. Here we recast the argument in the more modern language,  derive an explicit expression for the matrix element of the four-quark twist-four operator as a linear combination of the structure functions $F_1^{\gamma}$ and $F_{1}^{\gamma Z}$, and provide a corresponding formula for the shift in the asymmetry,  $R_1(\mathrm{HT})$. Moreover, the matrix elements of four-quark twist-four operators are known~\cite{Ellis:1982cd, Ji:1993ey} to satisfy the Callan-Gross relation. We exploit this property to show that $Y_1=1$ up to twist-four and that the twist-four contribution in the $Y_1$ term lies entirely in the factor $F_1^{\gamma Z}/F_1^\gamma$. This result implies that the $Y_1$ term in Eq.~(\ref{APVY1Y3-A})  is in principle a relatively clean probe of twist-four quark-quark correlations.

Following the notation of Ref.~\cite{Wolfenstein:1978rr}, we start with an isospin decomposition of the electromagnetic current and the vector part of the WNC as
\bea
J^\mu_\gamma &=& v_\mu + \frac{1}{3} s_\mu - \frac{1}{3}\lambda_\mu, \nn \\
J^{V\mu}_Z &=& 2\big [ (1-2 \sin^2\theta_W ) v_\mu  - \frac{2}{3}\sin^2\theta_W s_\mu -(\frac{1}{2} - \frac{2}{3}\sin^2 \theta_W) \lambda_\mu \big ],
\eea
where the isovector, isoscalar, and strange quark currents are, respectively,
\bea
v_\mu = \frac{1}{2}(\bar{u}\gamma_\mu u - \bar{d}\gamma_\mu d ), \qquad s_\mu = \frac{1}{2}(\bar{u}\gamma_\mu u +\bar{d}\gamma_\mu d ), \qquad \lambda_\mu = \bar{s}\gamma_\mu s 
\eea
and where we have omitted heavy quark contributions or simplicity (including them is straightforward). 
Using this isospin decomposition of the currents in the expressions for $W_{\mu \nu}^{\gamma}$ and $W_{\mu \nu}^{V;\gamma Z}$ given in Eqs.(\ref{W-def}) and (\ref{WVA-1}) respectively, we arrive at the following isospin decomposition for the hadronic tensors
\bea
\label{Wisospin}
W_{\mu \nu}^{\gamma} &=&  W_{\mu \nu}^{vv} + \frac{1}{9} W_{\mu \nu}^{ss} + \cdots ,\nn \\
W_{\mu \nu}^{V;\gamma Z} &=& 2(1-2 \sin^2\theta_W ) W_{\mu \nu}^{vv} - \frac{4}{9}\sin^2\theta_W W_{\mu \nu}^{ss} + \cdots ,
\eea
where the dots indicate contributions from strange and heavier quarks and the hadronic tensors $W_{\mu \nu}^{vv,ss}$ are defined as
\bea
\label{Wvvss}
W_{\mu \nu}^{vv} &=& \frac{1}{M} \sum_{X} (2\pi)^3 \delta^{(4)} (p_X - P -q) \langle X | v_{\mu} | D(P) \rangle ^*\langle X |v_\nu  | D(P) \rangle ,\nn \\
&=& \frac{1 }{2\pi M} \int d^4x \>e^{i q\cdot x} \langle D(P) | v_\mu (x) v_\nu (0) |D(P)\rangle, \nn \\
W_{\mu \nu}^{ss} &=& \frac{1}{M} \sum_{X} (2\pi)^3 \delta^{(4)} (p_X - P -q) \langle X | s_{\mu} | D(P) \rangle ^*\langle X |s_\nu  | D(P) \rangle ,\nn \\
&=& \frac{1}{2\pi M} \int d^4x \>e^{i q\cdot x} \langle D(P) | s_\mu (x) s_\nu (0) |D(P)\rangle. \nn \\
\eea
We ignore subleading contributions arising from the strange and heavier quarks in this analysis for simplicity. Contributions to the hadronic tensors involving a product of the isovector $v_\mu$ current with the isosinglet $s_\nu$ current vanish by isospin symmetry since the deuteron is an isoscalar state.  

Next we note that the difference of the $W_{\mu \nu}^{vv}$ and $W_{\mu \nu}^{ss}$ hadronic tensors is given by $W_{\mu \nu}^{du}$
\bea
\label{vv-elim}
W_{\mu \nu}^{du} &=& W_{\mu \nu}^{ss} - W_{\mu \nu}^{vv} , \nn \\
&=& \frac{1 }{2\pi M} \int d^4x \>e^{i q\cdot x} \langle D(P) | \frac{1}{2}\{ \bar{d}(x)\gamma_\mu d(x)\> \bar{u}(0)\gamma_\nu u(0) + (u\leftrightarrow d) \} |D(P)\rangle. \nn \\
\eea
As seen above, the operator in $W_{\mu \nu}^{du}$ is just ${\cal O}^{du}(x)$ of Eq.(\ref{Omunu}) which is manifestly a twist-four, four-quark operator involving the different up and down flavors of quark bilinears. In the context of the light cone operator product expansion (OPE), it contains no local operators involving only two quark or two gluon fields as occur at twist-two since the fields located at different positions along the light-cone have different flavor. 

The relation in Eq.~(\ref{vv-elim}) can be understood from the definitions of $W_{\mu \nu}^{vv}$ and $W_{\mu \nu}^{ss}$ given in Eq.~(\ref{Wvvss}) and noting that
\bea
v_\mu(x) v_\nu(0) - s_\mu(x) s_\nu(0) &=& - \frac{1}{2}\{ \bar{d}(x)\gamma_\mu d(x)\> \bar{u}(0)\gamma_\nu u(0) + (u\leftrightarrow d) \}.
\eea
We now define flavor-dependent structure functions $F_{1,2}^{vv,ss,du}$ corresponding to the hadronic tensors $W_{\mu \nu}^{vv,ss,du}$ as
\bea
\label{Wdu-1}
W_{\mu \nu}^{vv,ss,du} &=& \big ( -g_{\mu \nu} + \frac{q_\mu q_\nu}{q^2}\big)\frac{F_1^{vv,ss,du}}{M} + \big(P_\mu - \frac{P\cdot q}{q^2} q_\mu \big)\big (P_\nu - \frac{P\cdot q}{q^2}q_\nu \big)\frac{F_2^{vv,ss,du}}{ M P\cdot q}, \nn \\
\eea
so that from Eq.~(\ref{vv-elim}) we have the relation
\bea
\label{Fvv-elim}
F_{1,2}^{vv} &=& F_{1,2}^{ss} -F_{1,2}^{du},
\eea
which allows us to eliminate the $F_{1,2}^{vv}$ structure functions in favor of $F_{1,2}^{ss}$ and $F_{1,2}^{du}$.
The structure functions $F_{1,2}^{\gamma (\gamma Z)}$ can be related to the $F_{1,2}^{vv,ss,du}$ structure functions via Eqs.~(\ref{Wisospin}) and (\ref{Wdu-1}) as
\bea
\label{fdu-fss}
F_{1,2}^\gamma &=& F_{1,2}^{vv} + \frac{1}{9}  F_{1,2}^{ss} = \frac{10}{9}F_{1,2}^{ss} - F_{1,2}^{du}, \nn \\
F_{1,2}^{\gamma Z} &=& 2(1-\frac{20}{9} \sin^2\theta_W ) F_{1,2}^{ss}  -  2(1-2 \sin^2\theta_W ) F_{1,2}^{du},
\eea
where we have used Eq.~(\ref{Fvv-elim}) to eliminate $F_{1,2}^{vv}$ in favor of $F_{1,2}^{ss}$ and $F_{1,2}^{du}$. 

\subsection{Isolating twist-four contribution to the asymmetry}

Using Eq.~(\ref{fdu-fss}) for the structure functions $F_{1,2}^{\gamma(\gamma Z)}$ that appear in Eq.~(\ref{asym1}), the electron polarization asymmetry can be brought into the form
\bea
\label{asym2}
A_{RL} &=&- \Big ( \frac{G_F Q^2}{4\sqrt{2} \pi \alpha} \Big )  \frac{ 2g_A^e \Big (1-\frac{20}{9}\sin ^2 \theta_W \Big ){\cal F}^{ss} -2g_A^e \Big ( 1 -2 \sin^2\theta_W \Big ){\cal F}^{du} + g_V^e x (2-y)F_3^{\gamma Z} }{\frac{10}{9}{\cal F}^{ss}- {\cal F}^{du}}, \nn \\
\eea
where we have introduced the shorthand notation
\bea
\label{def1}
{\cal F}^{ss} &\equiv& 2xy F_1^{ss}- 2 \Big [1-1/y + \frac{x M}{ E} \Big ] F_2^{ss}, \nn \\
 {\cal F}^{du} &\equiv& 2xy F_1^{du} - 2 \Big [ 1 - 1/y + \frac{xM}{E}\Big ]F_2^{du}.
\eea
Next we note that the leading twist contribution to $F_1^{ss}$ and $F_2^{ss}$ satisfies the Callan-Gross relation so that
\bea
\label{CGrel}
F_{2;LT}^{ss} = 2xF_{1;LT}^{ss},
\eea
where the subscript $LT$ indicates that this relation generally holds only for the leading twist contributions.
It has also been shown~\cite{Ellis:1982cd, Ji:1993ey, Qiu:1988dn} that the four-quark twist-four contribution to $F_1^{du}$ and $F_2^{du}$ satisfies the Callan-Gross relation so that
\bea
\label{CGrel-du}
F_2^{du} = 2x F_1^{du}.
\eea
We outline an alternate derivation of this Callan-Gross relation for $F_{1,2}^{du}$ in Appendix \ref{cgdu}.

Equations (\ref{CGrel}) and (\ref{CGrel-du}) allow us to write
\bea
{\cal F}^{ss}_{LT} &=& 2x F_{1;LT}^{ss}\Big [ y  -2 + 2/y - \frac{2xM}{E}\Big ], \nn \\
 {\cal F}^{du} &=& 2x F_1^{du} \Big [ y  -2 + 2/y - \frac{2xM}{E}\Big ]\ \ \ ,
\eea
These relations allow us to write
\bea
\label{ratioduss}
\frac{{\cal F}^{du}}{{\cal F}^{ss}_{LT}} &=& \frac{F_1^{du}}{F^{ss}_{1;LT}},
\eea
Using Eqs.(\ref{def1}), (\ref{CGrel}),  (\ref{CGrel-du}), and (\ref{ratioduss}) the terms in Eq.~(\ref{asym2}) proportional to $g_A^e$ can be brought into the form
\bea
\label{apv-vv}
A_{RL}^{V} &=&- \Big ( \frac{G_F Q^2}{4\sqrt{2} \pi \alpha} \Big )  \frac{2 g_A^e \Big (1-\frac{20}{9}\sin ^2 \theta_W \Big )-2g_A^e \Big ( 1 -2 \sin^2\theta_W \Big )\frac{{\cal F}^{du}}{{\cal F}^{ss}_{LT}} }{\frac{10}{9}(1 -  \frac{9}{10}\frac{{\cal F}^{du}}{{\cal F}^{ss}_{LT}})}, \nn \\
&=&- \Big ( \frac{G_F Q^2}{2\sqrt{2} \pi \alpha} \Big ) \frac{ g_A^e \Big (1-\frac{20}{9}\sin ^2 \theta_W \Big )-g_A^e \Big ( 1 -2 \sin^2\theta_W \Big )\frac{F^{du}_1}{ F^{ss}_{1;LT}} }{\frac{10}{9}(1 -   \frac{9}{10}\frac{F^{du}_1}{F^{ss}_{1;LT}})}, \nn \\
&=& -\frac{9}{10} \Big ( \frac{G_F Q^2}{2\sqrt{2} \pi \alpha} \Big ) g_A^e\Big \{ \Big (1-\frac{20}{9}\sin ^2 \theta_W \Big )-\frac{1}{10} \frac{F^{du}_1}{ F^{ss}_{1;LT}} +\cdots \Big \}, \nn \\
\eea
where we have used the symbol $A_{RL}^V$ to denote the part of the asymmetry $A_{RL}$ proportional to $g_A^e$ that arises from an interference of the electromagnetic current  with the vector weak neutral current. The first equality in Eq.~(\ref{apv-vv}) is obtained by dividing the numerator and denominator of the terms proportional to $g_A^e$ in Eq.~(\ref{asym2}) by ${\cal F}^{ss}$ and using
\bea
\frac{{\cal F}^{du}}{{\cal F}^{ss}} &=& \frac{{\cal F}^{du}}{{\cal F}^{ss}_{LT}} +\text{subleading terms},
\eea
to make the replacement $\frac{{\cal F}^{du}}{{\cal F}^{ss}} \to \frac{{\cal F}^{du}}{{\cal F}^{ss}_{LT}} $. The subleading terms above denote contributions arising from the twist-four matrix element ${\cal F}^{du}$ multiplying subleading twist contributions to ${\cal F}^{ss}$. The second equality in Eq.~(\ref{apv-vv}) is obtained by using Eq.~(\ref{ratioduss}) and the last equality is obtained by expanding to linear order in the quantity $F_1^{du}/F_{1;LT}^{ss}$. The expression for $A^{RL}_V$ in Eq.~(\ref{apv-vv}) is just the sum of the leading twist Cahn-Gilman term and a twist-four contribution from a single four-quark matrix element $F_{1}^{du}$. Comparing to Eqs.~(\ref{CG-modify}) and (\ref{R-effects}) we obtain  the main result of this paper:
\be
\label{eq:R1HTgen}
R_1(\mathrm{HT})= -\frac{9}{10}\frac{1}{ (9-20\sin^2\theta_W)} \frac{F^{du}_1}{ F^{ss}_{1;LT}} \ \ \ .
\ee

We now derive expressions for $F_{1;LT}^{ss} $ and $F_1^{du}$ in terms of the $F_{1,2}^{\gamma (\gamma Z)}$ structure functions which will be useful for phenomenological analyses. The leading twist structure function $F_{1;LT}^{ss}$  can be related to the leading twist $F_{1;LT}^{\gamma (\gamma Z)}$ structure functions as
\bea
\label{fss-LT}
F^{ss}_{1;LT} &=& \frac{9}{10}F_{1;LT}^\gamma = \frac{1}{2(1-\frac{20}{9}\sin^2\theta_W)} F_{1;LT}^{\gamma Z},
\eea
which follows directly from Eq.~(\ref{fdu-fss}). At leading twist the structure function $F_{1;LT}^{\gamma}$ takes the well known form 
\bea
F_{1;LT}^\gamma (x) &=& \frac{1}{2} \sum_{q} e_q^2 (q_D(x) + \bar{q}_D(x)),
\eea
where $q_D(x)$ and $\bar{q}_D(x)$ denote the quark and anti-quark deuteron PDFs respectively. Treating the deuteron as an isoscalar combination of the proton and neutron so that we have the relation
\bea
q_D(x) = \frac{1}{2}(q_p(x) + q_n(x) ),
\eea
where $q_{p,n}(x)$ denote the proton and neutron PDFs respectively  and ignoring sea-quark contributions we get
\bea
\label{fss-2}
F_{1;LT}^\gamma (x)&=& \frac{1}{4} \Big [  e_u^2 \big [u_p(x) +u_n(x)\big ]+ e_d^2 \big [ d_p(x) + d_n(x) \big ] \Big ]\nn \\
 &=& \frac{5}{36}\big [u_p(x) +  d_p(x) \big ] = \frac{10}{9}F_{1;LT}^{ss},
\eea
where we have used the isospin relations $u_{p,n}(x)=d_{n,p}(x)$ and where the last equality follows from Eq.~(\ref{fss-LT}). Thus, the leading twist structure function $F_{1;LT}^{ss}$ can be simply expressed in terms of known proton PDFs. Substituting the resulting expression into Eq.~(\ref{eq:R1HTgen}) leads to the result 
\bea
\label{R1HT}
R_1(\mathrm{HT}) &=& \Bigg [\frac{-4}{5(1-\frac{20}{9}\sin^2 \theta_W)}\Bigg ]\frac{F_1^{du}}{u_p(x) + d_p(x)}. \nn \\
\eea

Next we turn to the four-quark twist-four contribution $F_{1}^{du}$. From Eq.~(\ref{fdu-fss}), $F_1^{du}$ is given in terms of the standard structure functions $F_1^\gamma$ and $F_1^{\gamma Z}$, which can be extracted from experiment, as 
\bea
\label{Fdu-3}
F^{du}_1 &=& \Big [ (9-20 \sin^2\theta_W) F_1^\gamma - 5 F_1^{\gamma Z} \Big ].
\eea
Note that the LHS of Eq.~(\ref{Fdu-3}) is manifestly a four-quark twist-four matrix element while the RHS includes twist-2 contributions, two-quark twist-four contributions and four-quark twist-four contributions. This implies that the twist-2 and two-quark twist-four contributions on the RHS cancel out. This allows us to write Eq.~(\ref{Fdu-3}) as
\bea
\label{Fdu-4}
F^{du}_1 &=& \Big [ (9-20 \sin^2\theta_W) F_1^{\gamma;4q} - 5 F_1^{\gamma Z;4q} \Big ], 
\eea
where the superscript $4q$ indicates that only the four-quark twist-four operator contributions to
$F_1^{\gamma (\gamma Z)}$ are kept. This makes the four-quark twist-four nature of the RHS in
Eq.~(\ref{Fdu-3}) manifest.

Using Eqs.(\ref{fss-2}) and (\ref{Fdu-4}) in Eq.~(\ref{eq:R1HTgen}), the twist-four contribution $R_1(HT)$ to the asymmetry from the interference of the electromagnetic current and the vector weak neutral current is given by
\bea
\label{shift-A}
R_1(\mathrm{HT}) &=&   -\frac{4}{5}\>\frac{ \Big [ (9-20 \sin^2\theta_W) F_1^{\gamma;4q} - 5 F_1^{\gamma Z;4q} \Big ]}{\Big (1-\frac{20}{9}\sin ^2 \theta_W \Big ) \Big [ u_p(x) +  d_p(x)\Big ]}.
\eea

\subsection{Equality of $R^\gamma$ and $R^{\gamma Z}$ at twist-four}
While the results in Eqs.~(\ref{eq:R1HTgen}) and (\ref{R1HT}) embody the observation of Refs. \cite{Bjorken:1978ry,Wolfenstein:1978rr} in the form of structure functions, the relationship to the parameterization of Eq.~(\ref{APVY1Y3-A}) for the asymmetry is not manifest. To make the implications for the latter apparent,  we draw on the analysis of the previous section to show that the relation
\bea
R^\gamma &=& R^{\gamma Z}
\eea
is valid at twist-four, implying that $Y_1=1$ up to perturbative corrections in $\alpha_s(Q^2)$. 
Using the following decomposition of the structure functions
\bea
\label{twist-decomp}
F_{1,2}^{\gamma (\gamma Z)} &=& F_{1,2; LT}^{\gamma (\gamma Z)}  + \delta F_{1,2}^{\gamma (\gamma Z)} ,
\eea
where $F_{1,2; LT}^{\gamma (\gamma Z)} $ denotes the leading twist contribution to $F_{1,2}^{\gamma (\gamma Z)}$ and $\delta F_{1,2}^{\gamma (\gamma Z)}$ denotes the higher twist contributions, we can write
\bea
\label{expand-twist}
R^{\gamma (\gamma Z)} &=& r^2 \frac{F_{2;LT}^{\gamma (\gamma Z)} + \delta F_2^{\gamma(\gamma Z)}}{2 x F_{1;LT}^{\gamma(\gamma Z)}} \big [ 1- \frac{\delta F_1^{\gamma(\gamma Z)}}{F_{1;LT}^{\gamma (\gamma Z)}}\big ] -1 + \cdots, \nn \\
&=& r^2 \big [1+ \frac{\delta F_2^{\gamma (\gamma Z)}-2x \delta F_1^{\gamma (\gamma Z)}}{2x F_{1;LT}^{\gamma (\gamma Z)}} \big ] -1 + \cdots,
\eea
where we have expanded $F_{1,2}^{\gamma (\gamma Z)}$ and the dots denote terms suppressed by higher powers of $Q^2$.  In what follows we only keep terms up to twist-four and suppress $ +\cdots $ terms in Eq.(\ref{expand-twist}). From this relation it follows that
\bea
\label{twist-decomp-1}
\frac{1+ R^{\gamma Z}}{1+R^\gamma} &=& 1 + \frac{\delta F_2^{\gamma Z}-2x \delta F_1^{\gamma Z}}{2x F_{1;LT}^{\gamma Z}} - \frac{\delta F_2^{\gamma }-2x \delta F_1^{\gamma}}{2x F_{1;LT}^{\gamma }}.
\eea
Using Eq.~(\ref{fdu-fss}) we can write
\bea
\label{twist-decomp-2}
F_{1,2;LT}^{\gamma} &=&  \frac{10}{9} F_{1,2;LT}^{ss}, \qquad F_{1,2;LT}^{\gamma Z} = 2(1-\frac{20}{9}\sin^2 \theta_W) F_{1,2;LT}^{ss}, \nn \\
\delta F_{1,2}^{\gamma} &=& \frac{10}{9}\delta F_{1,2}^{ss}  - F_{1,2}^{du}, \qquad \delta F_{1,2}^{\gamma Z} = 2(1- \frac{20}{9}\sin ^2 \theta_W)\delta F_{1,2}^{ss}  - 2(1-2\sin^2 \theta_W)F_{1,2}^{du},
\eea
where $\delta F_{1,2}^{ss}$ denotes the contribution to $F_{1,2}^{ss}$ from terms beyond twist-2. Using the expressions in Eq.~(\ref{twist-decomp-2}) in Eq.~(\ref{twist-decomp-1}) we arrive at
\bea
\frac{1+ R^{\gamma Z}}{1+R^\gamma} &=& 1 + \frac{(F_2^{du} - 2x F_1^{du})}{2x F_{1;LT}^{ss}} \Big [ \frac{9}{10} - \frac{1-2\sin^2\theta_W}{1-\frac{20}{9}\sin^2\theta_W}\Big ].
\eea
 Now using Eq.~(\ref{CGrel-du}) 
we arrive at the result
\bea
\frac{1+ R^{\gamma Z}}{1+R^\gamma} &=& 1,
\eea
valid up to twist-four, neglecting perturvative corrections in $\alpha_s(Q^2)$. This leads to the result
\bea
\label{Y1is1}
Y_1= 1,
\eea
up to corrections in $\alpha_s(Q^2)$ and power suppressed terms beyond twist-four.
Finally, using Eqs.(\ref{fdu-fss}) and (\ref{twist-decomp}), the ratio $F_1^{\gamma Z}/F_1^\gamma$ which appears in the $Y_1$ term of Eq.~(\ref{APVY1Y3-A}) can be written as
\bea
\label{F-ratio}
\frac{F_1^{\gamma Z}}{F_1^\gamma} &=& \frac{9}{5}\Big [ (1-\frac{20}{9}\sin^2 \theta_W) - \frac{1}{10}\frac{F_1^{du}}{F_{1;LT}^{ss}}\Big ],
\eea
leading immediately to Eq.~(\ref{apv-vv}).

\section{Model Estimates of Higher Twist Effects}
\label{sec:mit}

Given the unique sensitivity of the $Y_1$ term in the deuterium asymmetry to the four-quark HT operator $\mathcal{O}^{\mu\nu}_{ud}(x)$, it is useful to provide model estimates of its contribution as a benchmark for the Jefferson Lab PVDIS program. To that end, we utilize the MIT Bag Model\cite{Chodos:1974je}, following the analysis of Refs. ~\cite{Sacco:2004ck, Sacco:2009kj}. In principle, one may consider the use of other models to estimate the matrix element of $\mathcal{O}^{\mu\nu}_{ud}(x)$, such as QCD sum rules (see, {\em e.g.} Ref.~\cite{Stein:1995si} and references therein) or the instanton vacuum approximation\cite{Dressler:1999zi} that have been applied more extensively to HT effects in polarized structure functions. In addition, a nonperturbative QCD computation would yield a result from first principles. To our knowledge, the particular nucleon matrix element of interest here has not been computed in any of these approaches, though some indications may be inferred from related calculations. For example, the authors of Ref.~\cite{Dressler:1999zi} showed that in the instanton vacuum approximation, four quark matrix elements are suppressed relative to those of two-quark/gluon HT operators. 
The authors of Ref.~\cite{Capitani:1999rv} have carried out a quenched lattice computation of the contribution of the isospin two, four quark operator to the pion structure function using Wilson fermions, and find that its scale is set by the square of the pion decay constant, $F_\pi^2$. These authors suggested that the scale of the corresponding nucleon matrix elements would be set by $m_N$ rather than $F_\pi$, presumably leading to a larger value than in the constituent quark picture. In both cases, only contributions to the leading moments were considered.

In what follows, we will use the MIT Bag Model to estimate not only the overall magnitude of the HT four quark contribution but also its dependence on $x_B$. We note that the MIT Bag Model was previously employed by the authors of Refs.~\cite{Castorina:1985uw,Castorina:1984wd}. In applying their computation to the deuterium asymmetry, these authors assumed that the $x_B$-dependence of the twist-two and twist-four contributions to the structure functions were similar and obtained the twist-four contributions by rescaling the twist-two contributions by the ratio of their leading moments. Under these approximations, they obtained 
\be
\label{eq:mulders}
R_1(\mathrm{HT}) \approx -\frac{5.7\times 10^{-3}}{Q^2/(\mathrm{GeV})^2}\ \ \ .
\ee

\begin{figure}
\includegraphics[height=4in, width=6in]{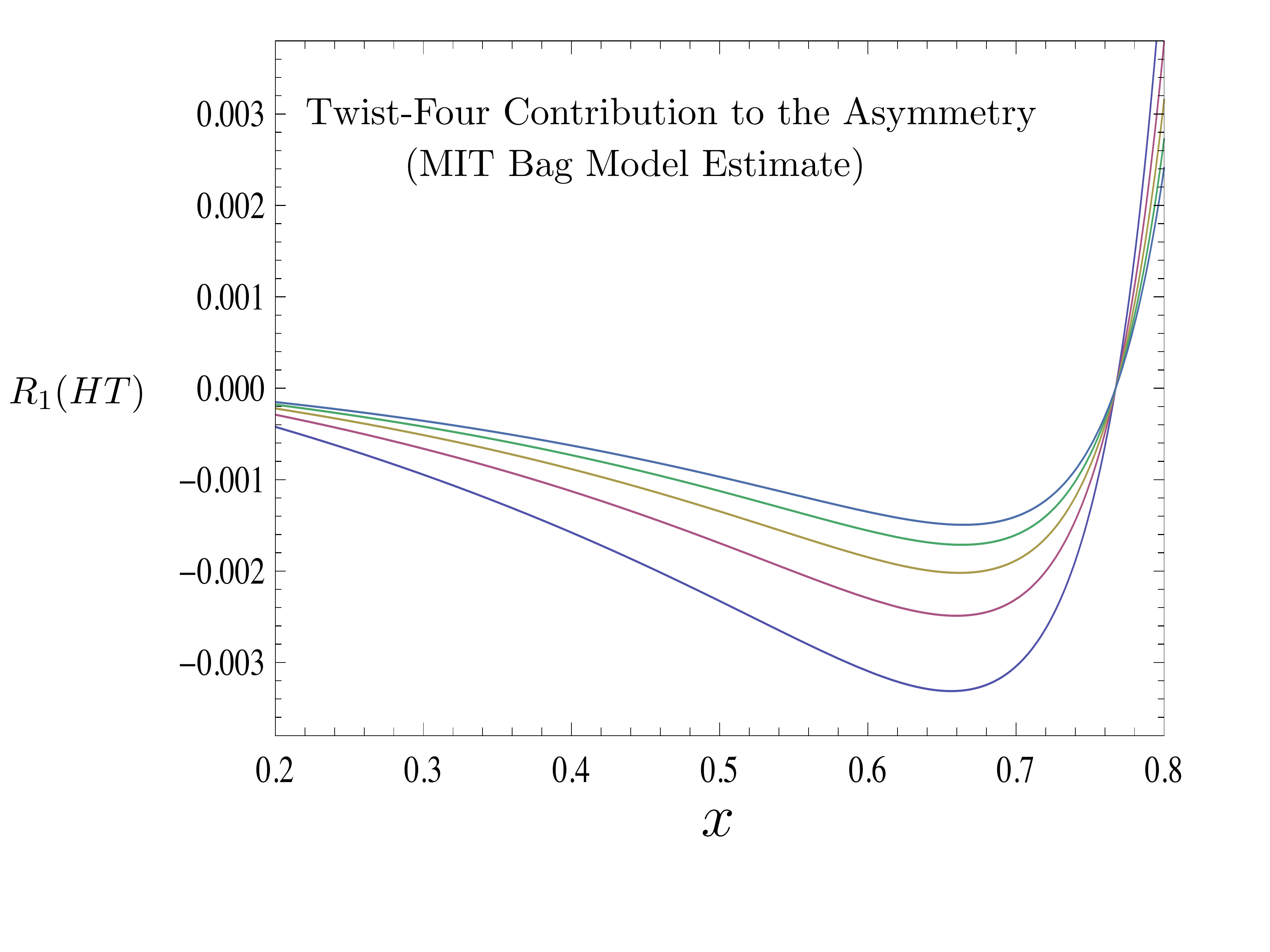}
\caption{The estimate of $R_1(HT)$ as a function of the Bjorken variable $x$ for different values of $Q^2$ in the MIT Bag Model. The curves from the bottom to top correspond to the values $Q^2=4,6,8,10,12$ GeV$^2$ respectively.}
\label{fig-asym}
\end{figure}

In what follows, we extend the analysis of Ref.~\cite{Castorina:1985uw} by allowing differences in the $x_B$-dependences of the twist-two and -four contributions to the structure functions. To this end, we compute a series of twist-four structure function moments, fit these moments to a parameterization in moment number $N$, and perform an inverse Mellin transform to obtain the structure function. This procedure is subject to several uncertainties, including the truncation of the tower of moments  before carrying out the inverse Mellin transform and the choice of parameterization to which they are fit. In addition, we have neglected the logarithmic evolution of the moments from the hadronic scale to the $Q^2$ of interest, as a full computation of the anomalous dimension matrix -- including the effects mixing between the four-quark and  twist-four quark-quark gluon and purely gluonic operators remains to be completed. Nonetheless, we believe the computation described below provides a reasonable model estimate for the four-quark structure function relevant to $R_1(\mathrm{HT})$. We find that the value of the structure function moments decreases rapidly with $N$, justifying the neglect of higher moments in the inversion,  and that our results for the inverse Mellin transforms do not vary appreciably as we change the parameterization used in fitting them. To the extent that the logarithmic $Q^2$ evolution of the moments at these scales is gentle and that the quark-quark correlations embodied in the MIT Bag Model capture the dominant twist-four physics, then the estimates described below and illustrated in Fig. \ref{fig-asym} should provide a reasonable benchmark.

The four-quark twist-four  contributions to $F_1^{\gamma (\gamma Z)}$ are given by\cite{Ji:1993ey,Ellis:1982cd}
\bea
\label{F1U1U2}
F_1^{\gamma;4q}(x_B,Q^2)&=& \frac{x_B}{2} \frac{\Lambda^2}{Q^2}\sum_{q,q'} \int dx dy dz \Big [ e_q e_{q'} U_1(x,y,z) \big (\Delta (x,y,z, x_B) + \Delta (y-x,y,y-z,x_B)  \nn \\
&-& \Delta(y-x,y,z,x_B) - \Delta(x,y,y-z,x_B)\big ) \nn \\
&+& e_q e_{q'} U_2(x,y,z) \big ( \Delta(x,y,z,x_B) + \Delta(y-x,y,y-z,x_B) \nn \\
&+& \Delta(y-x,y,z,x_B)+ \Delta(x,y,y-z,x_B)\big )\Big ], \nn \\
F_1^{\gamma Z;4q}(x_B,Q^2) &=& \frac{x_B}{2} \frac{\Lambda^2}{Q^2}\sum_{q,q'} \int dx dy dz \Big [ e_q g_{q'}^V U_1(x,y,z) \big (\Delta (x,y,z, x_B) + \Delta (y-x,y,y-z,x_B)  \nn \\
&-& \Delta(y-x,y,z,x_B) - \Delta(x,y,y-z,x_B)\big ) \nn \\
&+& e_q g_{q'}^V U_2(x,y,z) \big ( \Delta(x,y,z,x_B) + \Delta(y-x,y,y-z,x_B) \nn \\
&+& \Delta(y-x,y,z,x_B)+ \Delta(x,y,y-z,x_B)\big )\Big ], \nn \\
\eea
where $e_q$ and $g_q^V$ are, respectively, the quark electric charge and vector coupling to the $Z$-boson with $C_{1q}=-g_A^e g_q^V/2$ and
\bea
\Delta (x,y,z,x_B) &=& \frac{\delta(x-x_B)}{(y-x)(z-x)} + \frac{\delta(y-x_B)}{(x-y)(z-y)} + \frac{\delta(z-x_B)}{(y-z)(x-z)}\ \ \ ,
\eea
and the deuteron four-quark operator matrix elements $U_1$ and $U_2$ are given by
\bea
\label{eq:U1U2}
U_1(x,y,z) &=& \frac{g^2}{4 \Lambda^2} \int \frac{d\lambda}{2\pi}\frac{d\mu}{2\pi}\frac{d\nu}{2\pi} e^{i\lambda x}e^{i\mu(y-x)}e^{i\nu(z-y)} \langle D| \bar{\psi}^q(0) \nslash t^a \psi^q(\nu n)\bar{\psi}(\mu n) \nslash t^a \psi^{q'}(\lambda n) | D\rangle ,  \\
U_2 (x,y,z) &=& \frac{g^2}{4 \Lambda^2} \int \frac{d\lambda}{2\pi}\frac{d\mu}{2\pi}\frac{d\nu}{2\pi} e^{i\lambda x}e^{i\mu(y-x)}e^{i\nu(z-y)} \langle D| \bar{\psi}^q(0) \nslash \gamma_5 t^a \psi^q(\nu n)\bar{\psi}(\mu n) \nslash \gamma_5 t^a \psi^{q'}(\lambda n) | D\rangle , \nn \\
\eea
where $g$ is the $SU(3)_C$ color coupling constant and $t^a$ denote the generators of $SU(3)_C$ in the fundamental representation. We have introduced the hadronic scale $\Lambda$ in Eqs.~(\ref{F1U1U2},\ref{eq:U1U2}) in order to express the structure functions in terms of dimensionless functions, though the $F_1^{\gamma\ (\gamma Z)}$ are independent of this scale.
The  N-th moment of a structure function $F(x)$ is defined as 
\bea
M(N) &=& \int_0^1 dx x^{N-1} F(x),
\eea 
and the structure function can be obtained from its N-th moment via the inverse Mellin transform
\bea
F(x)&=& \frac{1}{2\pi i} \int_{c-i\infty}^{c+i\infty} dN x^{-N} M(N).
\eea
The second, fourth, and sixth moments of the $U_{1,2}$ contributions to $F_1^{\gamma (\gamma Z);4q}$ have been computed in the MIT Bag Model in ~\cite{Sacco:2009kj,Sacco:2004ck}. These results are given in Table. \ref{t4} and below we give a brief summary of the computation (for more of the moment calculation, see Refs.~\cite{Sacco:2009kj,Sacco:2004ck}).

The N-th moment of the structure function is parameterized as 
\bea
M(N) &=& \sum_{i=j}^{j+2} \frac{a_i}{N^i}.
\eea
After the inverse Mellin transform, the corresponding structure function takes the form
\bea
F(x) &=&   \sum_{i=2}^{4} a_i \frac{(-\log x)^{i-1}}{(i-1)!}.
\eea
The structure functions $F_1^{\gamma (\gamma Z);4q}$ receive contributions from two terms corresponding to the $U_1$ and $U_2$ contributions in Eq.(\ref{F1U1U2})
\bea
F_1^{\gamma;4q} &=& F_1^{\gamma;4q;U_1} +F_1^{\gamma;4q;U_2}, \nn \\
F_1^{\gamma Z;4q} &=&   F_1^{\gamma Z;4q;U_1} +F_1^{\gamma Z;4q;U_2}. \nn \\
\eea
For the numerical estimates we use the results of the fit for the $U_1$ and $U_2$ contributions at $Q^2=1$ GeV$^2$ which are given by
\bea
F_1^{\gamma; 4q; U1} &\simeq&  F_1^{\gamma Z; 4q; U1}: a_2 = 9.45 \times 10^{-4}, a_3= -277\times 10^{-4}, a_4 = 516.7 \times 10^{-4},\nn \\
F_1^{\gamma; 4q; U2} &\simeq&  F_1^{\gamma Z; 4q; U2}: a_2 = 164.5 \times 10^{-4}, a_3= -1197.0\times 10^{-4}, a_4 = 1191.4\times 10^{-4},\nn \\
\eea

\begin{table}
\begin{picture}(2,8)
\rput[tl](-10,8){
\centerline{\begin{tabular}{|c|c|}
\multicolumn{2}{c}{}\\
\hline
\multicolumn{2}{|c|}{$M(N)$ of $F_1^{\gamma Z;4q;U1}(x_B)$}\\
\hline
$M(2)$&$ 0$\\
$M(4)$ &$-1.72 \times 10^{-4}$\\
$M(6)$ & $-0.62 \times 10^{-4}$\\
\hline
\end{tabular}  }}
\rput[tl](-4,8){
\centerline{\begin{tabular}{|c|c|}
\multicolumn{2}{c}{}\\
\hline
\multicolumn{2}{|c|}{$M(N)$ of $F_1^{\gamma Z;4q;U2}(x)$ }\\
\hline
$M(2)$&$ -34.00 \times 10^{-4}$\\
$M(4)$ &$-3.77 \times 10^{-4}$\\
$M(6)$ & $-0.05 \times 10^{-4}$\\
\hline
\end{tabular}  }}
\rput[tl](-9.8,5){\centerline{\begin{tabular}{|c|c|}
\multicolumn{2}{c}{}\\
\hline
\multicolumn{2}{|c|}{$M(N)$ of $F_1^{\gamma ;4q;U1}(x_B)\>\>\>\>$}\\
\hline
$M(2)$&$ 0$\\
$M(4)$ &$-1.71 \times 10^{-4}$\\
$M(6)$ & $-0.61 \times 10^{-4}$\\
\hline
\end{tabular}  }}
\rput[tl](-4,5){\centerline{\begin{tabular}{|c|c|}
\multicolumn{2}{c}{}\\
\hline
\multicolumn{2}{|c|}{$M(N)$ of $F_1^{\gamma ;4q;U2}(x)$ }\\
\hline
$M(2)$&$ -33.72 \times 10^{-4}$\\
$M(4)$ &$-3.74 \times 10^{-4}$\\
$M(6)$ & $-0.05 \times 10^{-4}$\\
\hline
\end{tabular}  }}
\end{picture}
\vspace*{-2 cm}
\caption{The first few moments for the structure functions $F_1^{\gamma Z;4q;U1}(x), F_1^{\gamma Z;4q;U2}(x), F_1^{\gamma ;4q;U1}(x)$, and $F_1^{\gamma ;4q;U2}(x)$ in the MIT Bag model at $Q^2=1$ GeV$^2$. These are updated numbers after correcting some numerical errors discovered in \cite{Sacco:2004ck, Sacco:2009kj}.}
\label{t4}
 \end{table}
 

In order to determine the correction $R_1(\mathrm{HT})$, we substitute these results into the numerator of Eq.~(\ref{shift-A}). For the denominator we use the CTEQ5 pdfs \cite{Lai:1999wy}.
The resulting shift in the asymmetry, for different representative values of $Q^2$, is plotted in Fig.(\ref{fig-asym}). For the range of $x_B$ shown, the correction is largest in the valence region, reaching a magnitude commensurate with that obtained by the authors of Ref.~\cite{Castorina:1985uw}. The growth for $x_B$ near one is a result of the relatively quicker fall off with $x_B$ of the pdfs; the twist four contribution in the numerator is falling less quickly in this region\footnote{In general, the implementation of the parton model in the threshold region $x_B\to 1$ introduces theoretical ambiguities, so we avoid this kinematic regime in computing the relative correction. For a general discussion and references, see, {\em e.g.}, Ref.~\cite{Steffens:2006ds}.}. One can also perform this analysis with alternative parameterizations of the structure function moments. This was done in Ref. \cite{Sacco:2004ck} with similar results and we refer the reader to it for more details.

For the kinematic range that will be accessible in the planned JLab experiments, the Bag Model estimate for the magnitude of the correction $R_1(\mathrm{HT})$ lies below the expected experimental sensitivity. To the extent that the $Y_1$ term can be separated experimentally from the remaining contributions to the asymmetry, one could test this Bag Model expectation by looking for an appreciable $Q^2$ dependence in the data. The presence of such a dependence would point to stronger  correlations between quarks of different flavors than implied by the Bag Model picture, which correlates the up- and down-quarks largely through the confinement radius and the Pauli exclusion principle. On the other hand, the absence of large power corrections would imply that the $Y_1$ term can be interpreted primarily in terms of the underlying electroweak interactions and/or possible CSV in the parton distributions. We comment on the implications for probes of CSV and new physics in the following section.


%

\section{Charge Symmetry Violation and New Physics}
\label{sec:csv}
To the extent that $R_1(\mathrm{HT})$ is either tiny as suggested by the MIT Bag Model estimates or large enough to be extracted utilizing the $1/Q^2$-dependence, one may hope to use the deuterium asymmetry as a probe of CSV and/or new physics.  In terms of the former, it has recently been suggested that HT contributions to the $Y_1$ term in the deuterium asymmetry may be too large and too theoretically uncertain to utilize this term as a probe of CSV \cite{Hobbs:2008mm}. These suggestions were based on the possibility that $R^\gamma$ and $R^{\gamma Z}$ could differ substantially. We have shown that at finite $Q^2$ where a twist expansion is still valid, such
a possibility cannot apply at twist four since $R^\gamma = R^{\gamma Z}$ at this order in the twist expansion,  up to perturbative corrections. We now compare the MIT Bag Model estimate of $R_1(\mathrm{HT})$ to the CSV correction, $R_1(\mathrm{CSV})$. To that end, we follow the parameterization of CSV effects utilized in Ref.~\cite{Hobbs:2008mm}:
\bea
u_p & =&  u+ \frac{\delta u}{2}\nn \\
\label{eq:csv1}
d_p & =&  d+ \frac{\delta d}{2} \\
u_n & =&  d- \frac{\delta d}{2}\nn \\
d_n & =&  u- \frac{\delta u}{2}\ \ \ .\nn \\
\eea
In terms of the $\delta u$ and $\delta d$ one has
\be
\label{csv-param}
R_1(\mathrm{CSV}) = \left[ \frac{1}{2}\left(\frac{2C_{1u}+C_{1d}}{2C_{1u}-C_{1d}}\right)-\frac{3}{10}\right]\left(\frac{\delta u-\delta d}{u+d}\right)\ \ \ .
\ee
The $\delta u$ and $\delta d$ have been constrained by structure function data utilizing the {\em ansatz}
\bea
\label{kappa}
\delta u - \delta d &=& 2 \kappa f(x)\nn \\
f(x)&=& x^{-1/2}(1-x)^4(x-0.0909)\ \ \ ,
\eea
with $\kappa$ lying in the range $-0.8 \leq \kappa \leq +0.65$. Detailed phenomenological and theoretical analyses of CSV effects can be found in Refs.\cite{Martin:2003sk,Hobbs:2008mm, Londergan:2009kj}. In Fig. \ref{CSV}, we show the relative magnitudes of $R_1(\mathrm{HT})$ and $R_1(\mathrm{CSV})$ for a representative value of $Q^2=6\>$GeV$^2$ and $\kappa$ given by the extremes of the allowed range.  We observe that the Bag Model higher twist correction is considerably smaller than the possible range for CSV effects. To the extent that the Bag Model  provides a realistic guide for the magnitude of $R_1(\mathrm{HT})$, a series of precise measurements of the leading term in the asymmetry could provide a powerful probe of CSV effects.

The implications for probing new physics via $R_1(\mathrm{new})$ are less clear. To be concrete, we follow Ref.~\cite{RamseyMusolf:1999qk} and consider new parity-violating contact interactions
\be
\label{eq:lnew}
\mathcal{L}_\mathrm{new} = \frac{4\pi\kappa^2}{\Lambda^2}\ \bar{e}\gamma^\mu\gamma_5 e\ \sum_f\ h_V^f\ {\bar f}\gamma^\mu f\ \ \ ,
\ee
where $\Lambda$ is the mass scale associated with the new physics, $\kappa^2$ gives the overall coupling strength, and the $h_V^f$ are the specific vector current couplings to each fermion $f$. Retaining only contributions from up- and down-quarks, we obtain the corresponding contribution to the correction $R_1(\mathrm{new})$:
\be
\label{eq:r1new}
R_1(\mathrm{new}) = \left(\frac{-16\pi\kappa^2}{3}\right)\ \left(\frac{v}{\Lambda}\right)^2\ \left(\frac{2 h_V^u-h_V^d}{1-20\sin^2\theta_W/9}\right)\ \ \ ,
\ee
where we have expressed the Fermi constant in terms of the Higgs vaccum expectation value $v=246$ GeV. 

A given scenario for new physics will determine the specific values of $\kappa$, $\Lambda$, and the $h_V^f$. For example,  E$_6$ grand unified models contain additional U(1) gauge groups that may lead to  the existence of a TeV-scale $Z^\prime$ boson. To illustrate the sensitivity of the $Y_1$-term to this scenario, we consider a particular pattern of symmetry-breaking that gives rise to a low-mass $Z_\chi$ boson. In this case, the correction  $R_1(\mathrm{new})$ arises from tree-level exchange of the $Z_\chi$. In terms of the parameters appearing in Eqs.~(\ref{eq:lnew},\ref{eq:r1new}) one has $\kappa^2=2.2\alpha$, $\Lambda=M_\chi$, $h_V^u=0$, and $h_V^d=-1/20$. For $M_\chi=1$ TeV, we obtain  $R_1(\mathrm{new})=1.85\times 10^{-3}$ independent of $Q^2$. Comparing with Figures \ref{fig-asym} and \ref{CSV}, we observe that the scale of this correction is commensurate with that of $R_1(\mathrm{HT})$ in the MIT Bag Model and well below the allowed bands for the possible CSV correction. In order for a measurement of the $Y_1$-term to probe this scenario, one would need an experimental sensitivity of $\sim 0.2\%$ with knowledge of the CSV and HT corrections at a similar or better level theoretical precision. On the other hand, the planned 4\% measurement of the proton's weak charge by the Q-Weak experiment  at Jefferson Lab will probe the same scenario for a one TeV $Z_\chi$. A similar comparison with other scenarios suggests that for a determination of $R_1(\mathrm{new})$ to be competitive with the Q-Weak experiment as a probe of new physics\footnote{We note that the RHS of  Eq.~(45) of Ref.~\cite{RamseyMusolf:1999qk} contains an error and should be multiplied by a factor of eight. As a result, the mass bound scale factor for ${\tilde\delta}_1$ in Table I should be multiplied by $2\sqrt{2}$. The same factor should be applied to the last entries in Tables II-IV.}, one would need a combined experimental and theoretical uncertainty of better than $\sim 0.5\%$. At present, then, it appears that a study of the $Y_1$ term in the deuterium asymmetry is better suited as a probe of hadron structure than of new physics\footnote{We note that the possible contributions from supersymmeric extensions of the SM have been analyzed recently in Ref.~\cite{Kurylov:2003xa}, though the analysis applied to the asymmetry as a whole and not the $Y_1$ term alone.  After taking into considerations constraints from other electroweak precision observables and direct search limits, corrections of up to 1.5\% on the asymmetry are currently allowed in supersymmetric models. }. 

\begin{figure}
\includegraphics[width=6in, height=5in]{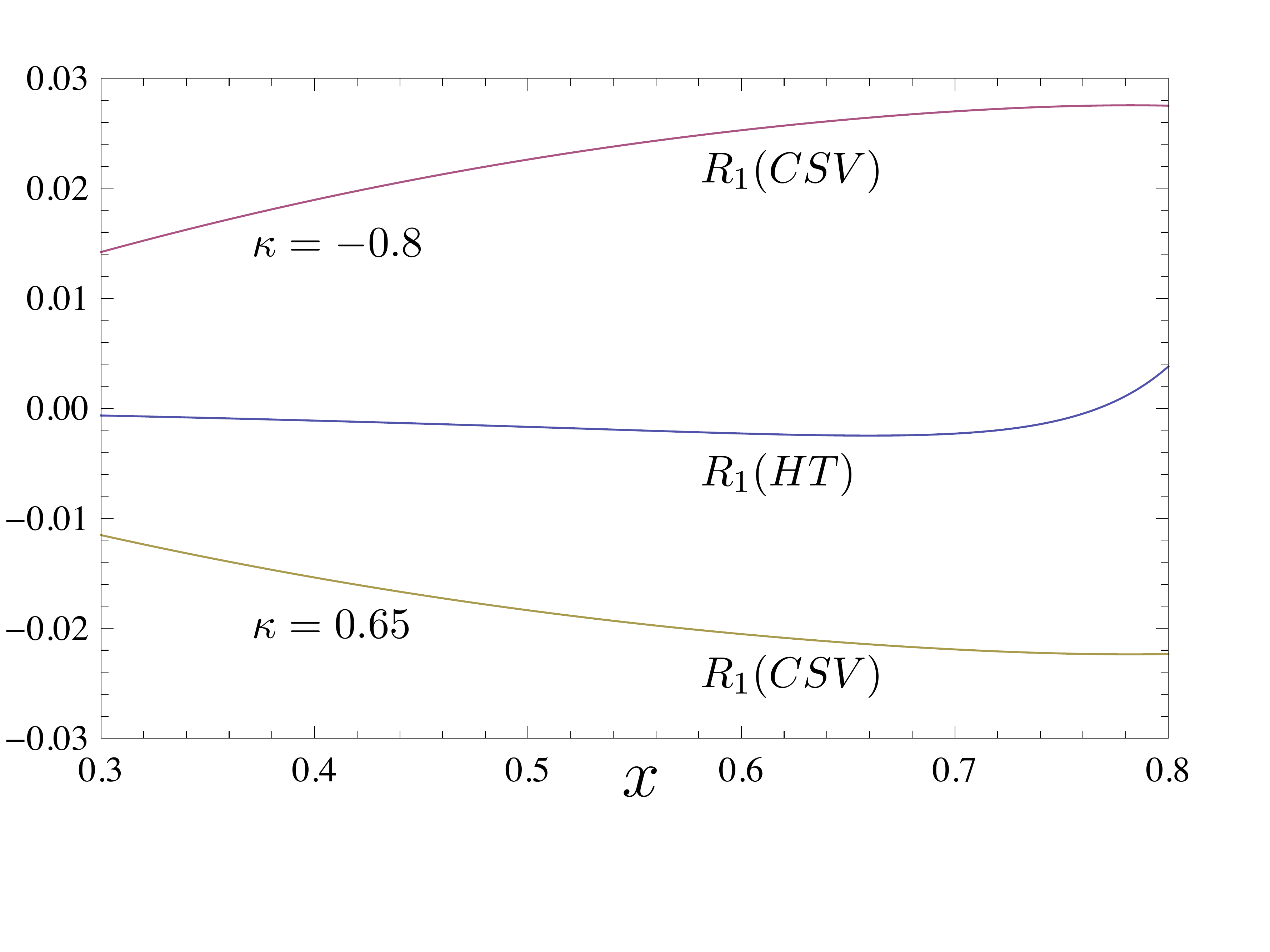}
\caption{The relative magnitudes of $R_1(HT)$ and $R_1(CSV)$ as a function of the Bjorken-$x$ variable for a representative value of $Q^2=6$ GeV$^2$. using $\delta u - \delta d = 2 \kappa f(x)$ where $f(x)= x^{-1/2}(1-x)^4(x-0.0909)$ for $\kappa =-0.8$. The top curve and bottom curves give $R_1(CSV)$ for the choices $\kappa=-0.8$ and $\kappa=0.65$ respectively in Eqs.(\ref{csv-param}) and (\ref{kappa}). The middle curve is the MIT Bag Model estimate for $R_1(HT)$. }
\label{CSV}
\end{figure}

\OMIT{
\begin{figure}
\includegraphics{CSV}
\rput[tl](-4.5,5.7){$R_1(CSV)$}
\rput[tl](-4.5,3.0){$R_1(HT)$}
\rput[tl](-4.5,1.2){$R_1(CSV)$}
\rput[tl](-5.2, 0.2){\Large{$x$}}
\rput[tl](-8.5,5.0){$\kappa=-0.8$}
\rput[tl](-8.5,1.6){$\kappa=0.65$}
\caption{The relative magnitudes of $R_1(HT)$ and $R_1(CSV)$ as a function of the Bjorken-$x$ variable for a representative value of $Q^2=6$ GeV$^2$. using $\delta u - \delta d = 2 \kappa f(x)$ where $f(x)= x^{-1/2}(1-x)^4(x-0.0909)$ for $\kappa =-0.8$. The top curve and bottom curves give $R_1(CSV)$ for the choices $\kappa=-0.8$ and $\kappa=0.65$ respectively in Eqs.(\ref{csv-param}) and (\ref{kappa}). The middle curve is the MIT Bag Model estimate for $R_1(HT)$. }
\label{CSV}
\end{figure}}

\section{Conclusions}
\label{sec:conclude}

Parity-violating electron scattering has become a powerful tool for probing both novel aspects of hadronic and nuclear structure as well as possible indirect signatures of physics beyond the Standard Model. Its efficacy depends on both significant experimental advances in controlling systematic uncertainties and attaining high statistics as well as on substantial  developments in the theoretical interpretation of the parity-violating asymmetries. PVDIS represents a prime example of this synergy between experiment and theory. The first measurements of the deep inelastic asymmetry for a deuterium target relied on the simplest parton-level description of hadrons, yet the result with a 17\% experimental uncertainty (for the two highest energy points) was sufficient to single out the Standard Model description of the weak neutral current interaction from other alternatives. Today, one anticipates lower-energy measurements at Jefferson Lab with experimental errors below one percent for individual kinematic points, making for $\mathcal{O}(0.5\%)$ combined uncertainties on quantities of interest. The challenge for theory is to provide a framework for interpreting such precise results. 

In this study, we have attempted to do so for the leading term in the deuterium asymmetry. In principle, it can be kinematically separated from the subleading term (suppressed by $1-4\sin^2\theta_W$), making it an object of interest in its own right. In going beyond the simplest parton model description of the deuteron structure and the Standard Model description of the weak neutral current interaction, one may expect  contributions to this term arising from higher twist operators, the violation of charge symmetry in the leading twist (parton model) terms, and new physics. 
Drawing on early work by Bjorken and Wolfenstein, who showed that this term depends on the matrix element of a single, non-local twist four operator, we have delineated the twist four contribution from those that may arise from CSV and new physics. In doing so, we have shown that that to this order in the twist expansion, one has $R^\gamma=R^{\gamma Z}$ up to perturbative corrections, making for a theoretically cleaner interpretation of the asymmetry than recently suggested in the literature. We have also utilized the MIT Bag Model to estimate the $x_B$-dependent twist-four correction and find that it is small compared to the range of possible CSV effects as implied by global fits to structure function data. Typical contributions from new physics are also smaller than the allowed CSV range. To the extent that the Bag Model provides a reasonable guide to higher twist quark-quark correlations, one would not expect to observe appreciable sub-leading power dependence on $Q^2$ in the leading term but would, on the other hand, be able to make a clean interpretation of this term in terms of CSV. On the other hand, experimental evidence for a substantial $Q^2$ power dependence would point to interesting non-perturbative dynamics underlying the higher-twist matrix elements. Either way, a determination of this leading term at the level of precision expected for the Jefferson Lab experiments would provide new insights into the behavior of non-perturbative QCD.


\begin{acknowledgments}
We thank A. Belitsky, C. Keppel, K. Kumar, T. Hobbs, T. Longergan, W. Melnitchouk,  P. Mulders, P. Reimer, P. Souder, and C. Weiss for helpful exchanges and references to the literature. This work was supported in part under  U.S. Department of Energy
contract DE-FG02-08ER4153, the Wisconsin Alumni Research Foundation, and the Alfred P. Sloan foundation. 
\end{acknowledgments}

\appendix

\section{Callan-Gross relation: $F_2^{du}=2xF_1^{du}$}
\label{cgdu}

It has been shown~\cite{Ellis:1982cd, Ji:1993ey,Qiu:1988dn} that the contribution from the four-quark twist-four operator in Eq.~(\ref{vv-elim}) to the structure functions $F_{1,2}^{\gamma (\gamma Z)}$ satisfies the Callan-Gross relation. Here we recast this argument using the language of the Soft-Collinear Effective Theory(SCET)~\cite{Bauer:2000yr, Bauer:2001yt, Bauer:2002nz}, which is an effective field theory  for describing the interactions of collinear and soft degrees of freedom and can be applied to electron-deuteron scattering in the Breit frame. In the language of SCET, the argument for the Callan-Gross relation becomes manifest via the structure of the leading order SCET operator that appears at twist-four from a tree-level matching.

Recall that the contribution of the twist-four four-quark operator to the hadronic tensor
\bea
\label{Wdu-def-appex}
W_{\mu \nu}^{du}
&=&\frac{1}{2\pi M} \int d^4x \>e^{i q\cdot x} \langle D(P) | \frac{1}{2}\{ \bar{d}\gamma_\mu d(x)\> \bar{u}\gamma_\nu u(0) + (u\leftrightarrow d)\} |D(P)\rangle, \nn \\
\eea
is parameterized in terms of the structure functions $F_{1,2}^{du}$ as
\bea
\label{Wdu-Appex}
W_{\mu \nu}^{du} &=& \big ( -g_{\mu \nu} + \frac{q_\mu q_\nu}{q^2}\big)\frac{F_1^{du}}{M} + \big(P_\mu - \frac{P\cdot q}{q^2} q_\mu \big)\big (P_\nu - \frac{P\cdot q}{q^2}q_\nu \big)\frac{F_2^{du}}{ M P\cdot q}, \nn \\
\eea
as first written in Eq.~(\ref{Wdu-1}). From this parameterization we note that the Lorentz invariant quantity $P^\mu P^\nu W_{\mu \nu}^{du}$ is given by
\bea
\label{FLproject}
P^\mu P^\nu W_{\mu \nu}^{du}
&=&\frac{(P\cdot q)^2}{q^2 M} \Big [ F_1^{du} - \frac{F_2^{du}}{2x}\Big ]. \nn \\
\eea
\begin{figure}
\includegraphics[height=1.5in, width=6in]{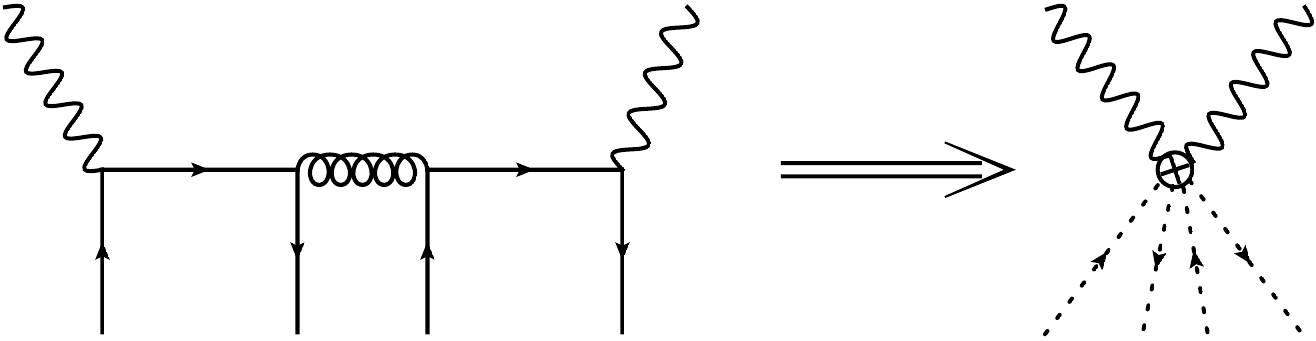}
\caption{Tree level matching of the operator ${\cal O}^{du}_{\mu \nu}$ onto the SCET operator in the Breit frame. The dashed fermion lines indicate collinear fields in the standard notation of SCET.}
\label{scet}
\end{figure}
If it can be shown that this quantity vanishes, it implies the Callan-Gross relation
\bea
\label{CGappex}
F_2^{du} &=& 2x F_1^{du}.
\eea
We formulate the argument in the Breit frame where the momentum of the virtual photon or Z-boson is
\bea
q^\mu &=& Q \frac{\bar{n}^\mu - n^\mu}{2},
\eea
and the momentum of the deuteron is
\bea
P^\mu &=& \bn \cdot P \frac{n^\mu}{2} +  \frac{M^2}{\bn \cdot P} \frac{\bn^\mu}{2},
\eea
where we have introduced the light-cone vectors
\bea
n^\mu &=& (1,0,0,1), \qquad \bn^\mu = (1,0,0,-1), \qquad n^2=\bn^2=0, \qquad \bn \cdot n =2.
\eea
From these relations and using momentum conservation in the Breit frame we have
\bea
\frac{n\cdot P}{\bn \cdot P} \simeq \frac{M^2}{Q^2} \ll 1,
\eea
so that the deuteron momentum is entirely along the light-cone $n^\mu$ up to power corrections in $M^2/Q^2$. 

In SCET in the Breit frame, at leading order in the power counting in $\Lambda_{QCD}^2/Q^2$ and at tree level, the four-quark operator in $W_{\mu \nu}^{du}$ will be matched onto an SCET operator~\cite{Marcantonini:2008qn}
\bea
\label{SCET-match}
 \bar{d}\gamma^\mu d(x)\> \bar{u}\gamma^\nu u(0) &=& n^\mu n^\nu \int d\omega_1d\omega_2d\omega_3d\omega_4 \> C(\omega_1,\omega_2,\omega_3,\omega_4) \nn \\
 &\times& \frac{1}{2}\{ (\bar{\xi}^d_n W)_{\omega_1} \frac{\bnslash}{2} (W^\dagger \xi^d_n)_{\omega_2} (\bar{\xi}^u_n W)_{\omega_3} \frac{\bnslash}{2} (W^\dagger \xi^u_n)_{\omega_4}+ (u\leftrightarrow d)\}, \nn \\
\eea
where we have used standard SCET notation, a detailed explanation of which, can be found in ~\cite{Bauer:2000yr, Bauer:2001yt, Bauer:2002nz}. This is schematically shown in Fig.~\ref{scet}. Here $C(\omega_1,\omega_2,\omega_3,\omega_4)$ denotes the matching Wilson coefficient.  $W$ denotes the momentum space Wilson lines
\bea
W = \Big [ \sum_{\text{perms}}\> \text{exp} \Big ( \frac{-g}{\bar{{\cal P}}} \bn \cdot A_{n,p}(x)\Big ) \Big ],
\eea
and the fields $A_{n,p}, \xi_{n,p}$ denote collinear gluon and quark fields which are fourier transformed to momentum space with respect to large light cone momentum component in the $n^\mu$ direction. The Wilson lines $W$, determined by collinear  gauge invariance in SCET, generate all the spin terms at twist-four.  The labels $p,\omega_i$ on the SCET fields denote the large part of the light cone momentum and the $x$ dependence of the collinear gluon and quark fields corresponds to residual momentum fluctuations. For more details of the matching and the SCET notation used here see~\cite{Marcantonini:2008qn}. What is relevant to our discussion is that the collinear quark fields $\xi_n$ that appear in the SCET operator satisfy the equation of motion
\bea
\nslash \xi_n = 0,
\eea
which is the reason that there are no terms proportional to $\bar{n}^\mu \bar{n}^\nu$ in the tree level matching in Eq.~(\ref{SCET-match}). Thus, contracting Eq.~(\ref{SCET-match}) with $n_\mu n_\nu$ gives zero using the property $n^2=0$. Applying  these considerations to Eq.~(\ref{Wdu-def-appex}) we get the relation
\bea
\label{Wdu-contract}
P^\mu P^\nu W_{\mu \nu}^{du}&\simeq & Q^2 n_\mu n_\nu W_{\mu \nu}^{du}, \nn \\
&=& 0, 
\eea
which from Eq.~(\ref{FLproject}) implies the tree level Callan-Gross relation of Eq.~(\ref{CGappex}) up to power corrections in $\Lambda_{QCD}^2/Q^2$.


\bibliographystyle{h-physrev3.bst}
\bibliography{pvdis.bib}

\begin{thebibliography}{10}

\bibitem{Prescott:1978tm}
C.~Y. Prescott {\em et~al.},
\newblock Phys. Lett. {\bf B77}, 347 (1978).

\bibitem{Prescott:1979dh}
C.~Y. Prescott {\em et~al.},
\newblock Phys. Lett. {\bf B84}, 524 (1979).

\bibitem{Cahn:1977uu}
R.~N. Cahn and F.~J. Gilman,
\newblock Phys. Rev. {\bf D17}, 1313 (1978).

\bibitem{Armstrong:2005hs}
G0, D.~S. Armstrong {\em et~al.},
\newblock Phys. Rev. Lett. {\bf 95}, 092001 (2005).

\bibitem{Acha:2006my}
HAPPEX, A.~Acha {\em et~al.},
\newblock Phys. Rev. Lett. {\bf 98}, 032301 (2007).

\bibitem{Aniol:2007zz}
HAPPEX, K.~Aniol,
\newblock Eur. Phys. J. {\bf A31}, 597 (2007).

\bibitem{:2009zu}
G0, D.~Androic {\em et~al.},
\newblock Phys. Rev. Lett. {\bf 104}, 012001 (2010).

\bibitem{Ito:2003mr}
SAMPLE, T.~M. Ito {\em et~al.},
\newblock Phys. Rev. Lett. {\bf 92}, 102003 (2004).

\bibitem{Spayde:2003nr}
SAMPLE, D.~T. Spayde {\em et~al.},
\newblock Phys. Lett. {\bf B583}, 79 (2004).

\bibitem{Heil:1989dz}
W.~Heil {\em et~al.},
\newblock Nucl. Phys. {\bf B327}, 1 (1989).

\bibitem{Baunack:2009gy}
S.~Baunack {\em et~al.},
\newblock Phys. Rev. Lett. {\bf 102}, 151803 (2009).

\bibitem{Anthony:2005pm}
SLAC E158, P.~L. Anthony {\em et~al.},
\newblock Phys. Rev. Lett. {\bf 95}, 081601 (2005).

\bibitem{VanOers:2007zz}
Qweak, W.~T.~H. Van~Oers,
\newblock Nucl. Phys. {\bf A790}, 81 (2007).

\bibitem{Souder:2008zz}
P.~Souder,
\newblock Prepared for 16th International Workshop on Deep Inelastic Scattering
  and Related Subjects (DIS 2008), London, England, 7-11 Apr 2008.

\bibitem{PVDIS:Jlab6}
X.~Zheng, P.~Reimer, and e.~a. R.~Michaels,
\newblock http://www.jlab.org/exp\_prog/proposals/08/PR-08-011.pdf.

\bibitem{PVDIS:JLab12}
P.~Reimer, X.~Zheng, and e.~a. K.~Paschke,
\newblock http://www.jlab.org/exp\_prog/proposals/07/PR12-07-102.pdf.

\bibitem{Bjorken:1978ry}
J.~D. Bjorken,
\newblock Phys. Rev. {\bf D18}, 3239 (1978).

\bibitem{Wolfenstein:1978rr}
L.~Wolfenstein,
\newblock Nucl. Phys. {\bf B146}, 477 (1978).

\bibitem{Fajfer:1984um}
S.~Fajfer and R.~J. Oakes,
\newblock Phys. Rev. {\bf D30}, 1585 (1984).

\bibitem{Castorina:1985uw}
P.~Castorina and P.~J. Mulders,
\newblock Phys. Rev. {\bf D31}, 2760 (1985).

\bibitem{Hobbs:2008mm}
T.~Hobbs and W.~Melnitchouk,
\newblock Phys. Rev. {\bf D77}, 114023 (2008).

\bibitem{Kumar:2010}
K.~Kumar,
\newblock private communication .

\bibitem{Chodos:1974je}
A.~Chodos, R.~L. Jaffe, K.~Johnson, C.~B. Thorn, and V.~F. Weisskopf,
\newblock Phys. Rev. {\bf D9}, 3471 (1974).

\bibitem{Ellis:1982cd}
R.~K. Ellis, W.~Furmanski, and R.~Petronzio,
\newblock Nucl. Phys. {\bf B212}, 29 (1983).

\bibitem{Ji:1993ey}
X.-D. Ji,
\newblock Nucl. Phys. {\bf B402}, 217 (1993).

\bibitem{Qiu:1988dn}
J.-W. Qiu,
\newblock Phys. Rev. {\bf D42}, 30 (1990).

\bibitem{Bauer:2000yr}
C.~W. Bauer, S.~Fleming, D.~Pirjol, and I.~W. Stewart,
\newblock Phys. Rev. {\bf D63}, 114020 (2001).

\bibitem{Bauer:2001yt}
C.~W. Bauer, D.~Pirjol, and I.~W. Stewart,
\newblock Phys. Rev. {\bf D65}, 054022 (2002).

\bibitem{Bauer:2002nz}
C.~W. Bauer, S.~Fleming, D.~Pirjol, I.~Z. Rothstein, and I.~W. Stewart,
\newblock Phys. Rev. {\bf D66}, 014017 (2002).

\bibitem{Musolf:1993tb}
M.~J. Musolf {\em et~al.},
\newblock Phys. Rept. {\bf 239}, 1 (1994).

\bibitem{Schienbein:2007gr}
I.~Schienbein {\em et~al.},
\newblock J. Phys. {\bf G35}, 053101 (2008).

\bibitem{Accardi:2008ne}
A.~Accardi and J.-W. Qiu,
\newblock JHEP {\bf 07}, 090 (2008).

\bibitem{Sacco:2004ck}
G.~F. Sacco,
\newblock UMI-31-66015.

\bibitem{Sacco:2009kj}
G.~F. Sacco,
\newblock (2009).

\bibitem{Stein:1995si}
E.~Stein, P.~Gornicki, L.~Mankiewicz, and A.~Schafer,
\newblock Phys. Lett. {\bf B353}, 107 (1995).

\bibitem{Dressler:1999zi}
B.~Dressler, M.~Maul, and C.~Weiss,
\newblock Nucl. Phys. {\bf B578}, 293 (2000).

\bibitem{Capitani:1999rv}
S.~Capitani {\em et~al.},
\newblock Nucl. Phys. {\bf B570}, 393 (2000).

\bibitem{Castorina:1984wd}
P.~Castorina and P.~J. Mulders,
\newblock Phys. Rev. {\bf D31}, 2753 (1985).

\bibitem{Lai:1999wy}
CTEQ, H.~L. Lai {\em et~al.},
\newblock Eur. Phys. J. {\bf C12}, 375 (2000).

\bibitem{Steffens:2006ds}
F.~M. Steffens and W.~Melnitchouk,
\newblock Phys. Rev. {\bf C73}, 055202 (2006).

\bibitem{Martin:2003sk}
A.~D. Martin, R.~G. Roberts, W.~J. Stirling, and R.~S. Thorne,
\newblock Eur. Phys. J. {\bf C35}, 325 (2004).

\bibitem{Londergan:2009kj}
J.~Londergan, J.~Peng, and A.~Thomas,
\newblock Rev.Mod.Phys. {\bf 82}, 2009 (2010).

\bibitem{RamseyMusolf:1999qk}
M.~J. Ramsey-Musolf,
\newblock Phys. Rev. {\bf C60}, 015501 (1999).

\bibitem{Kurylov:2003xa}
A.~Kurylov, M.~J. Ramsey-Musolf, and S.~Su,
\newblock Phys. Lett. {\bf B582}, 222 (2004).

\bibitem{Marcantonini:2008qn}
C.~Marcantonini and I.~W. Stewart,
\newblock Phys. Rev. {\bf D79}, 065028 (2009).

\end{thebibliography}

\end{document}